\documentstyle[preprint,aps,epsfig,amsmath,twocolumn]{revtex}

\tighten

\begin{document}

\preprint{\verb$Id: qtmndp.tex,v 1.27 1998/09/28 10:12:44 gagel Exp gagel $}
\draft

%\twocolumn[\hsize\textwidth\columnwidth\hsize\csname @twocolumnfalse\endcsname

\title{Quantum transport and momentum conserving dephasing}

\author{ Ivo Knittel, 
         Florian Gagel and
         Michael Schreiber }

\address{ 
          Institut f\"ur Physik, 
          Technische Universit\"at,
          D-09107 Chemnitz, Germany
        }

\date{\today}

\maketitle

\begin{abstract}
We study numerically the influence of momentum-conserving
dephasing on the transport in a disordered 
chain of scatterers.
Loss of phase memory is caused by coupling the transport channels
to dephasing reservoirs. In contrast to previously used
models, the dephasing reservoirs are linked to
the transport channels between the scatterers,
and momentum conserving dephasing can be investigated.  
Our setup provides
a model for nanosystems exhibiting conductance quantization 
at higher temperatures in spite of the presence of
phononic interaction. We are able to confirm numerically
some theoretical predictions.
\end{abstract}

\pacs{73.50.Bk, 73.20.Jc, 73.40.-c}

%73.50.Bk  General theory, scattering mechanisms
%73.40.-c  Electronic transport in interface structures
%73.20.Jc  Delocalization processes

%%%%%%%%%%%%%%%%%%%%%%%%%%%%%%%%%%%%%%%%%%%%%%%%%%%%%%%%%%%%%%%%%%%%%%%

\section{INTRODUCTION}

The recently discovered jumps in the conductance of 
metallic nanowires and nanocontacts 
\cite{Costa} have demonstrated impressively that
quantum effects can dominate transport properties  of small
structures even at ambient temperature.
This is generally explained by the lateral confinement which 
induces  a rather large subband spacing
of the transport modes of the conductor \cite{Beenakker}.
Within a single transport mode, phononic backscattering 
is considerably reduced because
of the large amount of momentum transfer of $p=2 \hbar k_F$
that has to be provided by a phonon.
However, there remains the question whether the
underlying picture of ``ideal conductors'' \cite{Landauer:70,Landauer:rest}
is appropriate to describe e.g.~metallic nanocontacts; if one
insisted in a description in terms of ideal waveguides, one
should account for the unavoidable imperfections of real systems
like impurities or geometric deviations from the ideal-waveguides setup.
The inclusion of coherent scattering
centers in corresponding transport models leads, however, to the almost 
immediate loss of quantization of the conductance in the absence of
some stabilizing mechanism. 

An interesting question is whether
dephasing, induced by phononic interaction, 
could provide such a  mechanism:
Interaction with longitudinal phonons, 
although not causing momentum transfer,
can well be at the origin of dephasing
which counteracts the weak localization of
coherent backscattering. In this way, temperature could
be helpful to re-establish quantization in this specific situation.
For an investigation of this question transport
models are needed which comprise elastic scattering 
as well as dephasing, while allowing for vanishing momentum
transfer.

Many approaches to dephasing
are based on reservoirs which destroy phase information
in the same way as momentum information of passing
electrons
\cite{Buettiker:86,Amato-pastawski,pastawski:91,McLDa:91,buschr:93};
therefore, corresponding transport models are not suited
for the investigation of the above discussed transport regime.

For the present study we have developed
a model where
coherent and incoherent scattering processes
are conceptually separated. Coherent scattering
is described by general elastic scatterers.
We investigate
single-mode transport through a one-dimensional chain
of scatterers (Fig.~\ref{fig1}). 

Dephasing 
is induced using a common
approach employing virtual 
electron reservoirs \cite{Buettiker:86}.
We have linked the reservoirs to the transport 
channels between elastic scatterers, exploiting
the fact that electrons 
in transport channels
automatically possess definite momenta. 
This seems to be the easiest way to 
implement momentum-selective coupling to a reservoir 
as has also been discussed by  Datta \cite{Datta1}.              
Chemical potentials of such reservoirs thus correspond 
to occupation numbers of a local momentum distribution function
as it occurs in the Boltzmann equation.
This implies furthermore that emission from the reservoirs
is uncorrelated with respect to the opposite direction of motion.

As depicted in Fig.~{\ref{fig1}}, electrons
absorbed by one of the two reservoirs between two
scatterers are fed back incoherently into either
the same transport channel or into the adjacent
channel. In the latter case, they conserve
their momentum as in the case of coherent
propagation, while having, however, lost their
phase memory.
This permits us to investigate continuously 
different degrees of momentum
conservation, from
the case of full momentum conservation to the
case of momentum flip, independently from the
dephasing strength itself.

Since
only two scattering channels per scatterer are
involved, our model is also efficient for
numerical calculations which is
of practical interest in order to cope
with ensemble-averaging.
It can also be easily extended 
towards multi-mode transport.

In the following section, we 
briefly generalize a method for the
calculation of the scattering matrix of a
composed system~\cite{gagelma:94} towards the
inclusion of dephasing reservoirs linked
to ``inner'' channels. Section III reviews the
involved characteristic lengths
for phase, momentum and coherent localization and
their relationship to the parameters of our model.
Numerical results for the metallic and insulating
regime are presented in section IV, and the influence
of momentum-conserving dephasing on the transport 
is discussed.

\section{Numerical approach}

\begin{figure}[ht]   %%%%%%%%%%%%%%%%%%%%%%%%%%%%%%%%%%%%%%% Figure 1
\begin{center}
\epsfig{file=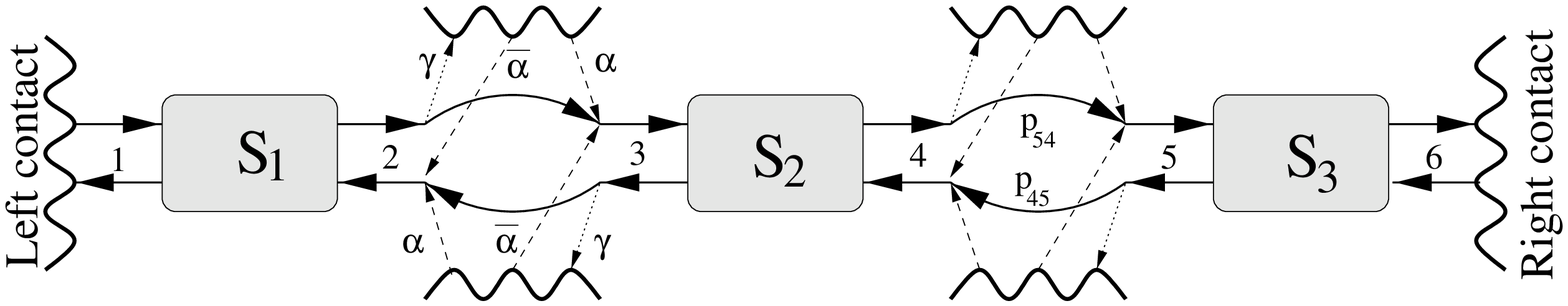,width=0.8\linewidth}\par
\vspace*{0.5cm}
  \caption{ \noindent
       A chain consisting of $N=3$ scatterers 
       $S_1$, $S_2$ and $S_3$. Channels
       $1$ and $6$ link the system to the left and right contact.
       Electrons outgoing from the scatterers 
       in inner channels 2--5 may travel
       coherently or incoherently between the scatterers.
       When traveling coherently, they aquire a phase shift
       $p$  (indicated by arcs).
       Alternatively, they may get absorbed (dotted arrows,
       probability $\gamma$) between the scatterers
       by reservoirs (wavy lines).
       From the reservoir they are re-emitted 
       (dashed  arrows) either with  same momentum
        (probability $\alpha$) or inversed momentum 
       (probability $\bar{\alpha}=1-\alpha$).
    \label{fig1}
  }
\end{center}
\end{figure}

We consider a one-dimensional chain consisting of $N$ scatterers
(see Fig.~\ref{fig1}). Each scatterer has two channels and is
described by its scattering matrix ${\bf S}_k$ which relates
linearly the outgoing to the incoming 
amplitudes at the scatterer.
We have used the parametrization
\begin{equation}
  {\bf S}_k = \left( \begin{array}{cc}
                      \sqrt{r_k}     &  \text{i} \sqrt{1-r_k} \\
                      \text{i} \sqrt{1-r_k} &    \sqrt{r_k}
                      \end{array}
              \right),
\end{equation}
where $r_k$ denotes the reflection probability of the
scatterer $k$, $k=1,\dots,N$.
A system of non-interacting scatterers would be described by
a $2N\times 2N$ scattering matrix $\bf K$ containing
the ${\bf S}_k$ on its diagonal.
Denoting the $2\,N$-tuples of incoming and outgoing amplitudes
by $c_+$ and $c_-$, respectively, we have $c_- = {\bf K} c_+$.
For linked scatterers, we may distinguish between
{\em external} and {\em internal} channels; external channels
are connected to the outside, e.g., the left and the right contact in
the present case (channels $1$ and $2N$), 
while internal channels are interconnected:
Amplitudes outgoing into internal channels are propagated
to incoming amplitudes in internal channels, thus acquiring
a phase shift $\exp(\text{i} p)$
which may comprise either a simple geometric phase
$p=q d$ for a  wave vector $q$ and a distance $d$, 
or, generally also a magnetic phase $\int {A} \cdot \text{d}{s}$ 
in presence of a vector potential ${A}$.  
This can be described by an operator $\bf{P}$
which propagates internally outgoing amplitudes while annihilating
externally outgoing amplitudes, 
$c_{+,\text{int}} = {\bf P} c_-$. Using
the decomposition $c_+ = c_{+,\text{int}} +c_{+,\text{ext}}$,
the relation
\begin{equation} \label{eq:mainsc}
  c_- = ({\bf K}^{-1} - {\bf P})^{-1} c_{+,\text{ext}}
\end{equation} 
is readily obtained \cite{gagelma:94}.
The wanted scattering matrix $\bf S$ of the composed system
relates incoming to outgoing amplitudes in external
channels, i.e.,
\begin{equation}
  c_{-,\text{ext}} \,=\, {\bf S} c_{+,\text{ext}};
\end{equation}  
it is thus given by
\begin{equation}
  {\bf S} = \left( 
        \left[
          ( {\bf K}^{-1}-{\bf P})^{-1}
        \right]_{i,j}
      \right), 
\end{equation}
where the indices $i$, $j$ label only the external channels.

The scattering matrix $\bf S$ then yields the transmission
matrix $\bf T$ according to $T_{ij} = |S_{ij}|^2$;
$T_{ij}$ is the transmission probability 
between the external channels $j$ and $i$.
The transmission matrix
thus relates tuples of (dimensionless) outgoing and
incoming currents,
\begin{equation}
  I_{-} \,=\, {\bf T} I_{+}.
\end{equation}
The associated incoming and outgoing
currents are given by the 
% $2\,N$-tupels of
the squared moduli of the amplitudes in the external channels.

To incorporate  dephasing in this approach, we assign
a virtual reservoir to each internal channel $l$
and  denote by $\gamma_l$ 
the probability of getting absorbed by such a reservoir.
The current into this reservoir is
thus given by $\gamma_l \,|c_-^{(l)}|^2$. Current conservation
then demands a corresponding loss term being accounted 
for also by the operator $\bf P$; therefore, the 
assigned phase factor
for coherent
propagation of $c_-^{(l)}$ has now to be multiplied
by $\sqrt{1-\gamma_l}$. For the considered
chain it is given by e.g.
\begin{equation}
  P_{ml} = \exp( \text{i} p_{ml} ) \sqrt{1-\gamma_l}
\end{equation}
where $m = l \pm 1$ for $l$ even or $l$ odd, respectively,
$p_{ml}$ denoting the phase for coherent propagation.
Speaking in terms of currents, only a fraction
$1-\gamma_l$ of the current outgoing in channel $l$
is propagated coherently while a fraction $\gamma_l$
is absorbed by the reservoir. 
In the following, we assume $\gamma_l \equiv \gamma$
throughout for all internal channels $l$ while we
can formally assign $\gamma_n=1$ to 
external channels $n$ which are fully
absorbing; in this way, the virtual reservoirs
are treated in the same way as the external contacts.
 
In the presence of the virtual reservoirs, the
dimension of the transmission matrix $\bf T$ has 
become larger and
the matrix elements describing transmission from
the virtual reservoirs are also needed.
They can be obtained by 
replacing $c_{+,\text{ext}}$ in  Eq.~\ref{eq:mainsc}
with $c_{+,\text{int}}\sqrt{\gamma}$ where  
$c_{+,\text{int}}$ is given by the $l$-th unit vector
for virtual reservoir $l$. 
The squared moduli of $c_-$ (multiplied by $\gamma$ for
internal channels $i$) then yield the elements $T_{il}$,
i.e., column $l$ of $\bf T$.

The virtual reservoirs can now be used to
introduce dephasing into the system \cite{Buettiker:86}.
For this purpose, electrons absorbed by the 
virtual reservoirs have to be 
re-emitted incoherently. The motion of electrons
inside the considered system (see Fig.~\ref{fig1})
can then be described as a Markovian process 
(see e.g.~\cite{pastawski:91}):
They are injected by the left contact, pass through 
 virtual reservoirs and finally get absorbed by either
the right or the left contact. Adopting this point of
view, we may renounce the notation of a chemical potential
\cite{Buettiker:86} which is convenient since we are interested
in the total transmission only.
 
It is at the level of the transmission matrix where
current conservation is imposed.
We describe momentum conservation
by a parameter $\alpha$ ranging from $0$ to $1$;
electrons absorbed by a reservoir between
two scatterers in the chain conserve their momentum
with probability $\alpha$ while their momentum
gets flipped with probability 
$\bar{\alpha} = 1-\alpha$ (see Fig.~\ref{fig1}).
The total transmission $T$ is then given by
\begin{eqnarray} T = &&\\ \nonumber
  T_{2N,1} &+& \sum_{l=1}^{N-1} \left[
                      (  T_{2N,2l+1} \alpha +
                         T_{2N,2l} \bar{\alpha} )
                         T_{2l,1} \right. \\ \nonumber
               &+&
                      \left. ( T_{2N,2l} \alpha +
                         T_{2N,2l+1} \bar{\alpha})  T_{2l+1,1}  
                        \right] \\ \nonumber 
                &+& \sum_{l,l'=1}^{N-1} \left( \dots \right) + \dots
\end{eqnarray}
as sum of transmission probabilities from the
left to the right contact with $0,\; 1,\; \dots$
intermediate passages through virtual reservoirs. 
This infinite series describing the random walk
of the electrons can be summed analytically.
We note that $\bf T$ itself is symmetric 
due to time reversal invariance.
Therefore, it is no surprise that
the virtual reservoirs flip the momenta
of absorbed electrons if the latter are fed back into the
same channel; incoming and outgoing amplitudes in
the same channel correspond to opposite momenta.
Thus, our procedure of imposing current conservation only
onto a pair of channels is well beyond the 
widely used description of B\"uttiker\cite{Buettiker:86}; 
it is, however, covered
by the more general approach of Datta \cite{Datta1}.  

% \acknowledgements 

\section{Characteristic lengths}
 
The phase coherence length $l_\Phi$ is given by the mean
distance that an electron travels coherently, i.e.,
the distance it conserves its phase.
In order to relate it to the absorption probability $\gamma$,
we consider the unperturbed system (without disorder,
i.e., vanishing elastic backscattering) 
and obtain $l_\Phi$ in units of the number of passed dephasing regions
between the scatterers as
\begin{equation}
  l_\Phi = \frac{ \sum_{n=1}^{\infty} n (1-\gamma)^n }
                { \sum_{n=1}^{\infty} (1-\gamma)^n   } = \frac{1}{\gamma}.
\end{equation}  
The mean free path $l_m$ is the mean distance that
an electron travels without being scattered, i.e.,
without momentum change.
In order to obtain $l_m$ 
in the unperturbed system, 
we consider the (incoherent) 
reflection probability $\tilde{r}$ of a single dephasing
region which is  given by
$ \tilde{r} = \bar{\alpha} \gamma$.
The resulting (dimensionless) resistance caused by a 
pair of corresponding reservoirs is 
$\rho_0 = \tilde{r}/(1-\tilde{r})$, which is additive for two incoherenty coupled scatterers \cite{Buettiker:86}. 
We now show that $l_m := \rho_0^{-1}$ is a useful
definition for the mean free path:
For a chain of $N$ dephasing areas,
this yields a serial (four-probe) resistance of
$ \rho =\, N l_m^{-1}$,  
and, because of $\rho= (1-T)/T$,
a total transmission of
\begin{equation} \label{tlm}
  T = \frac{1}{1 + N / l_m}. 
\end{equation}
Equation \ref{tlm} describes an Ohmic length
dependence of the transmission;
a section of length $l_m$ contributes a resistance 
unit to the total resistance \cite{Datta1}.

The effect of the coherent 
disorder can be described by the localization
length $l_\xi$. We are extracting
this length from the ensemble-averaged transmission of purely
coherent systems, using the quantum scaling law of 
Anderson et~al.~(Ref.~\cite{pwa}), 
\begin{equation} \label{eq:explaw}
 \rho(N) = e^{N/l_\xi}-1
\end{equation}  
which differs from the well-known 
Landauer result \cite{Landauer:70} by a factor of 2.
Ref.~\cite{pwa} predicts, assuming random phases between the scatterers,
a localization length of the ensemble, measured in number of scatterers, as
\begin{equation} \label{eq:lxidef}
  l_\xi = \langle t \rangle / \langle r \rangle 
\end{equation}
with the ensemble averages of the reflection and transmission
probabilities $r$ and $t=1-r$ of a single scatterer. 
For our numerical investigation, we have
employed a constant reflection probability
$r_k \equiv r$ and random 
phases $p_{2k+1,2k}=p_{2k,2k+1} \in [0,2\pi[$
(indicated by arcs in Fig.~\ref{fig1}).

\begin{figure}[ht]   %%%%%%%%%%%%%%%%%%%%%%%%%%%%%%%%%%%%%%% Figure 2
\begin{center}
\epsfig{file=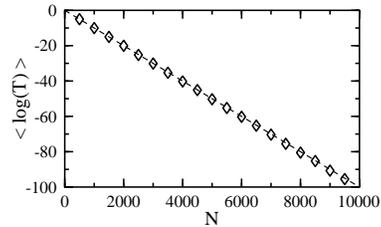,width=0.7\linewidth}\par
\vspace*{0.5cm}
  \caption{ \noindent
       Transmission of a chain of $N$ 
       coherently coupled scatterers
       with backscattering probability $r=0.01$. 
       Each point is obtained by logarithmically averaging
       $10^4$ realizations. The dashed line corresponds
       to Eq.~\ref{eq:explaw} with a localization length 
        $l_\xi = 99 = (1-r)/r$.
    \label{fig2}
  }
\end{center}
\end{figure}

Performing numerical tests
by logarithmically averaging up to $10^4$ realizations
of a purely
coherent system ($\gamma =0$)  
consisting of up to $N=5\times 10^4$ 
scatterers, we have a found quite a good agreement 
with Eqs.~\ref{eq:explaw}, \ref{eq:lxidef} 
(see Fig.~\ref{fig2}).
We have also tried a random
box distribution with average $r$ instead of
constant reflection probability;
this
neither had an influence
on the validity of Eq.~\ref{eq:lxidef}, nor on 
the following results involving both
coherent and incoherent processes.\\

\section{Results}

In order to investigate the interplay between the 
``microscopic'' parameters $r$, $\gamma$ and $\alpha$ and 
their counterparts, the
characteristic lengths $l_\xi$, $l_\Phi$ and $l_m$, 
we distinguish between the 
``conducting'' regime
$l_{\Phi}<l_{\xi}$ and the
``insulating'' regime
where $l_{\xi}<l_{\Phi}$.

Two representative results are shown in
Fig.~\ref{fig3} (left and right, respectively), 
where the computed ensemble-averaged total transmission
is plotted as a function of the number of scatterers
for dephasing which conserves, randomizes and
flips momenta ($\alpha=1$, $0.5$ and $0$, respectively). 
Transmission of the fully coherent system ($\gamma=0$) is similar to 
Fig.~\ref{fig2} (but not exactly the same due to the smaller number of averaged realizations).

The ``conductor'' can be thought of being composed
of phase-coherent units which are linked incoherently, 
resulting in an Ohmic length dependence of transmission. 
Since dephasing destroys 
localization, dephasing with full momentum conservation
($\alpha=1$) always enhances transmission when compared 
to a fully coherent system ($\gamma=0$ in Fig.~\ref{fig3}).

The widely discussed case of simultaneous phase and momentum 
randomization\cite{Buettiker:86} 
corresponds to $\alpha=0.5$; here, with
increasing number of scatterers, 
transmission first drops due to incoherent backscattering,
but there is always a certain length $N$ such that
the enhancement of transmission 
due to suppression of coherent localization
outperforms the inelastic backscattering for systems larger than
$N$. 
The absolute value of the transmission, however,
already has become very small. Therefore,
enhanced conductance  due to dephasing
can only be expected for a large degree of momentum
conservation $\alpha$.

In the insulating regime, the dephasing rate, but also the backscattering 
rate are much smaller than in above discussed, ``conducting'' case.
With $l_\xi<l_\Phi$, conduction properties are dominated by localization
effects and cannot be understood in
terms of incoherently coupled phase coherent units.
Dephasing supports transmission significantly only for system
sizes larger than $l_\Phi$ when a deviation from the
exponential dependence (described by Eq.~\ref{eq:lxidef}) 
in the right diagram of Fig.~\ref{fig3} can be noticed.
Finally, one would also expect an Ohmic behaviour 
for very large system sizes $N \gg l_\Phi$. Some enhancement of transmission
 in the insulating regime compared to the conducting regime appears but only 
for $\alpha<1$ (which can be seen comparing the two plots in Fig.~3).
This is not unphysical: A large $l_\Phi$ means small dephasing, which leads to a suppression of the
quantum localization of the coherent system and thus increases transmission.
For even larger dephasing (approaching the conducting regime), we get an 
increasing additional Ohmic resistance which finally causes the transmission to drop again (for $\alpha<1$).   

In the insulator, the relative enhancement of the transmission
for $\alpha=1$ compared to $\alpha=0.5$ or $\alpha=0$ 
is much smaller than for the case of the conductor,
momentum conservation has less influence in this regime.

The importance of momentum conservation is best seen
in  Fig.~\ref{fig4} where the dependence of the
transmission on the dephasing rate $\gamma$ is plotted.
Only for $\alpha=1$, transmission increases  monotonously
with increasing $\gamma$. In the incoherent
limit $\gamma=1$, an Ohmic transport regime is reached
with each single scatterer contributing the resistance $l_\xi^{-1}$
(see Eqs.~\ref{tlm}, \ref{eq:lxidef}), resulting in a transmission
\begin{equation}
   T=1/(1+N/l_m+N/l_\xi).
\label{tlm2}\end{equation}
For $\alpha=0.5$ and $\alpha=0$, transmission passes 
through a certain maximum which has already
been found for the case $\alpha=0.5$ 
in Ref.~\cite{Buettiker:86};
beyond this maximum, the additional resistance of the
dephasing reservoirs causes a rapid decline.

In the following, we compare
our numerical results (Figs.~\ref{fig3} and \ref{fig4}) 
to a formula which has been 
proposed by Band et~al.\cite{band} 
on the basis of ensemble averaged quantities.
The dashed lines in  Figs.~\ref{fig3} and \ref{fig4}
have been calculated using a slightly modified
version of this formula.

In this approach, the transmission is given as
\begin{equation}  
 T = \frac{1}{1 + N/l_m + \rho_d(l_\xi, l_\Phi,N)}\label{band}
\end{equation}
with $\rho_d$ being 
the resistance due to disorder for the case of infinite $l_m$.
Equation \ref{band} can be seen as a generalization of Eq.~\ref{tlm}
towards the inclusion of disorder.
The validity of Mathiessen's law is assumed
in the sense that the two processes which cause momentum change,
namely momentum flip by the reservoirs and coherent reflection 
by the scatterers, give independent contributions
to the resistance. Note that they do not depend 
on $l_\xi$ and $\alpha$, respectively.
Surely, quantum resistance is qenerally not additive. However, the range of
validity of the approach of Ref.~13 is surprisingly large (see Fig.~4).
Our numerical verification of the result of Band et al.~ shows 
excellent agreement. A pre-requisite for this is our dephasing model,
which allows for momentum conservation. 

In the presence of dephasing (for finite $l_\Phi^{-1}$),
$\rho_d$ can be built up recursively from the resistance of the 
purely coherent case, 
$\rho_c(x;l_\xi)=e^{x/l_\xi}-1$, and 
from the probability density 
$P(x; N, l_\xi, l_\Phi)$ for an electron to 
travel coherently a distance $x$ within the disordered wire 
and to lose phase information at
position $x$ by a dephasing event.  
Both $\rho_c(x;l_\xi)$ and $P(x; N, l_\xi, l_\Phi )$ are 
ensemble averages.
In the case $l_{\Phi}<N$, 
Band et~al.~obtain  
\begin{equation} \label{band3}
  \rho_d( N, l_\xi,l_\Phi) = \frac{N\int_0^N P(x; N, l_\xi,l_\Phi)
           \rho_c(x;l_{\xi}){\rm d}x}
            { \int_0^N x P(x; N, l_{\xi}, l_{\Phi}) 
              {\rm d}x}.
\end{equation}
Since both localization and dephasing 
processes lead to an exponential length decay of
the probability, 
\begin{equation}
  P(x; N, l_\xi,l_\Phi) = 
 l_{\lambda}^{-1}e^{-x/l_{\lambda}} +
  e^{-N/l_{\lambda}}\delta(x-N),
 \label{band2}
\end{equation}
with $x\in[0,N]$, is a plausible choice, with a disorder 
modified phase coherence length $l_{\lambda}(l_\xi,l_\Phi)$. 
In their original work, Band et~al.~have proposed 
\begin{equation}
  l_{\lambda}^{-1}= l_{\Phi}^{-1}+2 l_\xi^{-1};
\end{equation}
however, in more recent work \cite{Schr,Hey},
a phase coherence length in a 
disordered quantum wire is defined in a similar context 
as 
\begin{equation} \label{eq:schrey}
  l_{\lambda}^{-1} = l_{\Phi}^{-1}+l_\xi^{-1}.
\end{equation} 
Our numerical results are better described 
using Eq.~\ref{eq:schrey}.

Equation \ref{band} has been 
employed for Figs.~\ref{fig3} and \ref{fig4}.
This gives a rather good agreement, except for
the case $l_\Phi > l_\xi$ (right diagram of
Fig.~\ref{fig3}) where the contribution of inelastic
backscattering ($\alpha=0$) to the resistance
is slighly underestimated; Mathiessen's law
may thus not be valid in this regime.

For the interesting case of enhanced dephasing rates,
Eq.~\ref{band} can be used to
estimate the achievable change of transmission 
compared to purely coherent transmission.
The maximum enhancement $\Delta T$ 
is obtained for $l_\Phi=0, l_m=\infty$;
using Eqs.~\ref{eq:explaw} and \ref{tlm2}, we obtain
with $\chi:=N/l_\xi$
 \begin{equation}
    \Delta T = 1/(1+\chi)-e^{-\chi}.
 \end{equation}
This yields a maximum $\Delta T_{\text{max}} \approx 0.20$ for
$\chi_{\text{max}} \approx 2.51$.

\section{Conclusions}
\begin{figure}[ht]   %%%%%%%%%%%%%%%%%%%%%%%%%%%%%%%%%%%%%%% Figure 3
\hspace*{-1cm}
\epsfig{file=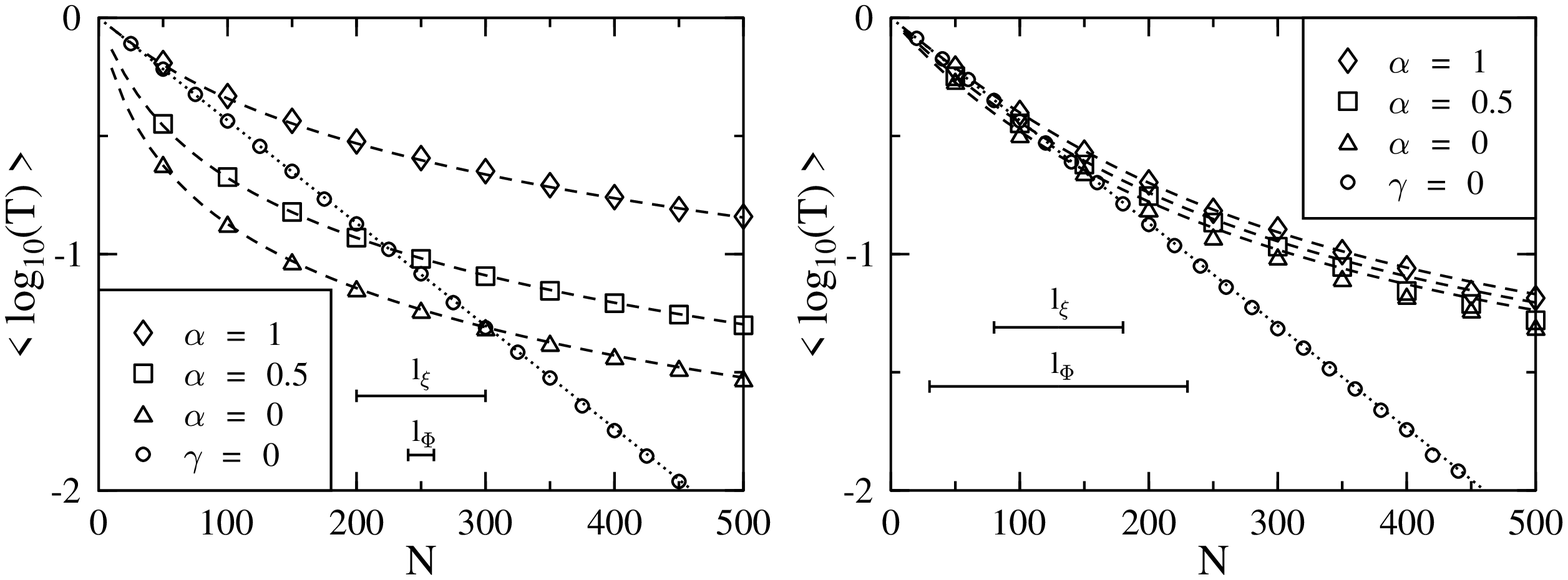, width=1.1\linewidth} 
\begin{center}
\vspace*{0.5cm}
  \caption{ \noindent
       Transmission as function
       of the number of scatterers for three different degrees
       of momentum conservation $\alpha$. The elastic backscattering
       probability is $r=0.01$ corresponding to a localization
       length $l_\xi=99$. The left diagram is calculated 
       for dephasing probability $\gamma=0.05$,
       the right diagram for $\gamma=0.005$.
       Each point is obtained by logarithmically averaging
       1000 realizations. Also indicated is the transmission
       of the coherent system ($\gamma=0$).
       The dashed and dotted lines indicate 
       solutions of Eq.~\ref{band}
       and Eq.~\ref{eq:lxidef}, respectively.
    \label{fig3}
  }
\end{center}
\end{figure}

\begin{figure}[ht]   %%%%%%%%%%%%%%%%%%%%%%%%%%%%%%%%%%%%%%% Figure 4
\epsfig{file=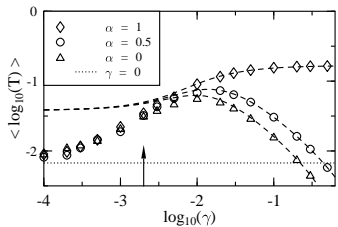, width=0.6\linewidth}
\begin{center}
\vspace*{0.5cm}
  \caption{ \noindent
       Total transmission for a chain of $N=500$ scatterers
       as a function of the dephasing parameter $\gamma$
       for three different degrees
       of momentum conservation $\alpha$. The elastic backscattering
       probability is $r=0.01$ corresponding to a localization
       length $l_\xi=99$;
       each point is obtained by logarithmically averaging
       1000 realizations. The lines are obtained from
       Eq.~\ref{band} which is applicable
       for $\gamma > 1/N$ (indicated by the arrow).
    \label{fig4}
  }
\end{center}
\end{figure}

We have presented a model of general scatterers 
which allows us to 
investigate quantum transport through a disordered system
in the  presence of dephasing with a tunable degree
of momentum conservation.
We obtained the rather obvious result
that momentum conservation is essential
if dephasing is to enhance 
transmission. From Fig.~\ref{fig3},
it is evident that support for a quantized conductance
by dephasing can be expected
in the regime $l_\Phi < N \approx l_\xi$.

Our numerical results have been compared 
to a general but heuristic picture
developed in Ref.~\cite{band} which is based
on the validity of Mathiessen's rule and an
ad hoc assumption for a probability 
density $P(x;N, l_{\xi}, l_{\Phi})$;
it is not derived from a microscopic 
theory, but gained its plausibility from its
validity in various limiting cases. 
Furthermore, it is not evident that 
average quantities as $\rho_c(x)$ 
and $P(x;N, l_{\xi}, l_{\Phi})$ can be employed 
in a recursive formula for the transmission, 
since sample averaging is not a simple commutative operation
and since there are different types of averaging procedures 
which often proved intuition wrong in the past \cite{pwa}.
Therefore, it is  rather surprising that 
Eq.~\ref{band} gives a good fit of our data, 
except for the strongly 
insulating regime, $l_{\xi} \ll l_{\Phi} \ll N$.

However, while the previously discussed approach by Band 
et~al.~yields a good agreement  for the ensemble-averaged 
transmission, it cannot predict fluctuations.
For the investigation of conductance quantization
effects, it is highly desirable to describe
a single system or at least the applicability
of results for ensemble averages to a 
particular realization. 
Here, our general quasi-microscopic model provides a helpful tool. 
In nanosystems exhibiting conductance quantization, elastic backscattering is reduced (but not absent) because of the larger subband spacing which is itself
induced through the geometrical confinement. At higher temperatures, one has
to account for phononic interaction, notably with logitudinal phonons, leading
to dephasing but not inelastic backscattering. As mentioned in the introduction, one can describe this situation by momentum-conserving dephasing which was not included in previous models. While we found an excellent agreement with the approach of Ref.~13 over a wide range, the domain of applicability to
(quasi-)quantization in nanosystems would be for $N\le l_\Phi< l_\xi$. Work along these lines is
in progress. An advantage compared to an heuristic approach is that we are able to investigate also fluctuations and single realizations.

%%%%%%%%%%%%%%%%%%%%% References  %%%%%%%%%%%%%%%%%%%%%%%%%%%%%

%%%%%%%%%%%%%%%%%%%%%%%%%%%%%%% Figures %%%%%%%%%%%%%%%%%%%%%%%%%%%%%%

\begin{references} 

\bibitem{Costa}J.~L.~Costa-Kr\"amer, N.~Garc\'ia, H.~Olin,
                   Phys.~Rev.~Lett.~{\bf 78}, 4990 (1997);
              J.~L.~Costa-Kr\"amer, N.~Garc\'ia, P.~Garc\'ia-Mochales,
               P.~A.~Serena, Surf.~Sci.~Lett.~{\bf 342}, L1144 (1995)


\bibitem{Beenakker} H.~van Houten, C.~Beenakker, Physics Today {\bf 7}
                    (July),
                    22 (1996) % ``7'' maybe not correct, July instead.     

\bibitem{Landauer:70} R.~Landauer, Philos.~Mag.~{\bf 21}, 863 (1970)

\bibitem{Landauer:rest} R.~Landauer, Z.~Phys.~B {\bf 68}, 217 (1987);
              R.~Landauer, J.~Phys.: Condens.~Matter {\bf 1}, 8099  (1989)


%\bibitem{Gagelmaqh}F.~Gagel, K.~Maschke,
%                   Phys.~Rev.~B {\bf 52}, 2013 (1995);
%                   F.~Gagel, M.~Schreiber, 
%                   K.~Maschke, Superlatt.~Microstruct.,
%                   {\bf 23}, No.~3/4, 593, (1998).


\bibitem{Buettiker:86}M.~B\"uttiker, Phys.~Rev.~B {\bf 33}, 3020 (1986)
        %% equivalence of virtual voltage probes and dissipation

\bibitem{Amato-pastawski} J.~L.~D'Amato and H.~M.~Pastawski, Phys.~Rev.~B 
{\bf 41}, 7411 (1990)


\bibitem{pastawski:91}H.~M.~Pastawski, Phys.~Rev.~B {\bf 44}, 6329 (1991)

\bibitem{McLDa:91} M.~J.~McLennan, Y.~Lee, S.~Datta,
                    Phys.~Rev.~B {\bf 43}, 13846 (1991)
%% Voltage drop in mesoscopic systems: A numerical study using
%% a quantum kinetic equation: ein mega paper, dissipation inside.



%% Classical and quantum transport from generalized L.-B. equations
%% Green's functions, Keldish-formalism




\bibitem{buschr:93} G.~Burmeister, K.~Maschke, M.~Schreiber, 
                   Phys.~Rev.~B {\bf 47}, 7095 (1993)          


\bibitem{Datta1} S.~Datta,``Electronic transport in mesoscopic systems'', 
                 Cambridge University Press, Cambridge 1993


\bibitem{gagelma:94}F.~Gagel, K.~Maschke,
                   Phys.~Rev.~B {\bf 49}, 17170 (1994)



\bibitem{pwa}P.~W.~Anderson, D.~J.~Thouless, E.~Abrahams, D.~S.~Fisher,
             Phys.~Rev.~B {\bf 22}, 3519 (1980)
% scaling rho = exp(a L)-1


\bibitem{band}Y.~B.~Band, H.~U.~Baranger, Y.~Avishai,
              Phys.~Rev.~B {\bf 45}, 1488 (1992)


%\bibitem{maschkeschreiber:91} K.~Maschke, M.~Schreiber,
%                      Phys.~Rev.~B {\bf 44}, 3835 (1991)

%\bibitem{maschkeschreiber:94} K.~Maschke, M.~Schreiber,
%                   Phys.~Rev.~B {\bf 49}, 2295 (1994)



%\bibitem{hershfielddephasing:91} S.~Hershfield, Phys.~Rev.~B {\bf 43}, 
%                          11586 (1991)
%% Equivalence of the multilead approach to dephasing and the
%% self-consistent Borm approximation


%\bibitem{onsager:44}L.~Onsager, Phys.~Rev.~{\bf 65}, 117 (1944)




\bibitem{Schr} K.~Maschke, M.~Schreiber, 
               Phys.~Rev.~B {\bf 49}, 2295 (1994)
\bibitem{Hey} R.~Hey, K.~Maschke, M.~Schreiber, 
             Phys.~Rev.~B {\bf 52}, 8184 (1995)


%%%%%%%%%%%%%%%%%%%%%%%%% Datta macht dasselbe 
%%\bibitem{datta} S.~Datta, Phys.~Rev.~B {\bf 45}, 1347 (1992)



\end{references}
\end{document}